\documentclass{ws-ijmpe}

\usepackage{dcolumn,array,bm,amsmath}

\newcommand{\beq}{\begin{equation}}
\newcommand{\eeq}{\end{equation}}
\newcommand{\ba}{\begin{array}}
\newcommand{\ea}{\end{array}}
\newcommand{\bn}{\begin{eqnarray}}
\newcommand{\en}{\end{eqnarray}}
\newcommand{\bbox}[1]{\bm{#1}}
\newcommand{\thalf}{\tfrac{1}{2}}
\newcommand{\tquart}{\tfrac{1}{4}}
\begin{document}

\markboth{S.G. Rohozi\'nski, J. Dobaczewski, W. Nazarewicz}{
Spatial symmetries of the local densities}

\catchline{}{}{}{}{}
%
%%%%%%%%%%%%%%%%%%%%%%%%%%%%%%%%%%%%%%%%%%%%%%%%%%%%%%%%%%%%%%%%%%%%

\title{SPATIAL SYMMETRIES OF THE LOCAL DENSITIES
%\footnote{For the title, try not to
%use more than 3 lines. Typeset the title in 10 pt
%Times roman, uppercase and boldface.}
}

\author{\footnotesize STANIS{\L}AW G. ROHOZI\'NSKI
%\footnote{Typeset names in
%10~pt Times roman, uppercase. Use the footnote to indicate
%the present or permanent address of the author.}
}

\address{Institute of Theoretical Physics, University of Warsaw, ul. Ho\.za 69\\
PL-00681 Warsaw, Poland
%\footnote{State completely without abbreviations, the
%affiliation and mailing address, including country. Typeset in 8~pt
%Times italic.}
\\
Stanislaw-G.Rohozinski@fuw.edu.pl}

\author{JACEK DOBACZEWSKI}

\address{
Institute of Theoretical Physics,
University of Warsaw,
ul. Ho\.za 69\\
PL-00681, Warsaw, Poland \\
Department of Physics, University of Jyv\"askyl\"a, P.O. Box 35 (YFL)\\
FI-40014 Jyv\"askyl\"a, Finland\\
Jacek.Dobaczewski@fuw.edu.pl}

\author{WITOLD NAZAREWICZ}
\address
{Department of Physics and Astronomy,
The University of Tennessee\\
Knoxville, Tennessee 37996, USA\\
Physics Division,
Oak Ridge National Laboratory,
P.O. Box 2008\\
Oak Ridge, Tennessee 37831, USA\\
Institute of Theoretical Physics,
University of Warsaw,
ul. Ho\.za 69\\
 PL-00681, Warsaw, Poland\\
witek@utk.edu}

\maketitle

\begin{history}
\received{(received date)}
\revised{(revised date)}
%\accepted{(Day Month Year)}
%\comby{(xxxxxxxxxx)}
\end{history}
%Wersja poprawiona po przyjêciu do druku
\begin{abstract}
Spatial symmetries of the densities appearing in the nuclear Density
Functional Theory are discussed. General forms of the local densities are derived by using methods of
construction of isotropic tensor fields.  The spherical and axial cases are considered.
\end{abstract}

\section{Introduction}\label{intr}

The nuclear Density Functional Theory (DFT) has become a conventional
approach to describing complex nuclei.\cite{Ben03,Lal04,Sto06} In
nuclear theory, the nuclear DFT is closely linked with the Hartree-Fock-Bogoliubov (HFB)
approximation to the nuclear many-body problem formulated in the
position space. The mean value of the system's energy
can here be expressed as  a functional of  the
single-particle density matrices in  the
particle-hole (p-h)  and particle-particle (p-p or pairing)
channels. The density matrices in both channels are expressed through
the scalar-isoscalar, scalar-isovector, vector-isoscalar and
vector-isovector densities, which are nonlocal, i.e., they depend on two
position vectors. In the local theory, e.g., the local density approximation, the nonlocal densities can be
represented in the energy density through local densities and their
derivatives.

The nuclear self-consistent mean field resulting from
the DFT may spontaneously break the symmetry of the original
many-body Hamiltonian. The symmetry-breaking phenomenon
allows for capturing important  physics effects and for
including essential correlations in the
many-body system. The existence of a self-consistent symmetry (SCS),
i.e., a symmetry obeyed by the mean-field Hamiltonian, reflects
physical properties of the system. Due to the self-consistency, the
SCS of the mean field is also the symmetry of the density
matrix. This  has a number of  important
consequences. First of all, to understand SCSs of DFT,  it is sufficient to
study symmetries of the density matrix without considering the
specific form of the density functional and resulting mean-field
Hamiltonian. We also realize that if the initial density matrix used
at the first iteration of the self-consistent procedure has a certain SCS, then this
symmetry will propagate through to the final DFT solution. This means
that the choice of the initial density matrix is crucial for the
proper physical content of the DFT results. Sometimes, e.g.,
 to simplify numerical calculations, a SCS of the
initial density matrix is {\it ad hoc} assumed. Such an assumption
may lead to overestimation of
the DFT energy and incorrect prediction of many properties of the system. Finally, assuming
a SCS, one should know the general form of the density matrix possessing
that symmetry.

Some point symmetries and the associated symmetry-breaking schemes of
the p-h densities have been investigated in Refs.\cite{Dob00a,Dob00b}
A detailed analysis of  SCSs of the DFT densities in both channels has recently been
presented in Ref.\cite{Roh09}, which  will be referred to as I in the following.
The present paper is a comment to I, with the main goal of
detailing the
consequences of spatial symmetries discussed therein. Although our
discussion  remains rigorous, in this contribution  we
simplify the notation
so as to make the results more transparent and understandable.

As in I, we treat the nonlocal and local densities as isotropic tensor
fields, i.e., functions of the position vector(s) independent of other
tensor quantities (material tensors).\cite{Tru52}
Under this  assumption, in Refs.\cite{Roh09,Pro09} we introduced the
Generalized Cayley-Hamilton (GCH) theorem for tensor fields,
and we used it as a tool to study general forms of local
densities. Since many different generalizations of the Cayley-Hamilton
theorem exist in the literature, in Sec.~\ref{GCHT}
we give a brief explanation of the present version.

In Sec.~\ref{den} we recall definitions of the density
matrices and  nonlocal and local densities. The spherical symmetry with
and without space inversion is discussed in Sec.~\ref{sph}. In
Sec.~\ref{axial}, we consider the axial symmetry with and without space
reflection. The summary of our study is contained in Sec.~\ref{sum}.

\section{Density matrices and densities}\label{den}
Within the HFB approach to the nuclear many-body problem, the mean
value of a many-body Hamiltonian is a functional (the energy
functional) of the p-h and p-p density matrices defined,
respectively, as
\begin{eqnarray}
\label{rho}
\hat{\rho}(\bbox{r}st,\bbox{r}'s't')
&=&\langle \Psi |a_{\bbox{r}'s't'}^{+}a_{\bbox{r}st}|\Psi \rangle , \\
\label{rhobreve}
\hat{\breve{\rho}}(\bbox{r}st,\bbox{r}'s't')
&=& 4s't'\langle \Psi |a_{\bbox{r}'-s'-t'}a_{\bbox{r}st}|\Psi \rangle ,
\end{eqnarray}
where $a_{\bbox{r}st}^{+}$ and $a_{\bbox{r}st}$ create and
annihilate, respectively, nucleons at point $\bbox{r}$, spin
$s$=$\pm\thalf$, and isospin $t$=$\pm\thalf$, while $|\Psi \rangle $
is the HFB independent-quasiparticle state. The properties of density matrices that
directly result from their definitions are the following:
\bn
\hat{\rho}^{\ast}(\bbox{r}st,\bbox{r}'s't')&=&\hat{\rho}(\bbox{r}'s't',\bbox{r}st), \label{herm}\\
\hat{\breve{\rho}}(\bbox{r}st,\bbox{r}'s't')&=&-16ss'tt'\hat{\breve{\rho}}(\bbox{r}'-s'-t',\bbox{r}-s-t). \label{antys}
\en
The dependence on the spin and isospin variables in the density
matrices can be easily separated by expanding in the spin and isospin
Pauli matrices $\hat{\bbox{\sigma}}_{ss'}$ and $\hat{\tau}^k_{tt'},\
(k=1,\, 2,\, 3)$, respectively:
\bn
\hat{\rho}(\bbox{r}st,\bbox{r}'s't')
&=&\tquart\left(\rho_0(\bbox{r},\bbox{r}')\delta_{ss'}
+\bbox{s}_0(\bbox{r},\bbox{r}')\cdot \hat{\bbox{\sigma}}_{ss'}\right)\delta_{tt'}\nonumber \\
&+& \tquart\sum_k\left(\delta_{ss'}{\rho}_k(\bbox{r},\bbox{r}')
+ \bbox{s}_k(\bbox{r},\bbox{r}')\cdot \hat{\bbox{\sigma}}_{ss'}\right) \hat{\tau}_{tt'}^k,
\label{izo01} \\
\hat{\breve{\rho}}(\bbox{r}st,\bbox{r}'s't')
&=&\tquart\left(\breve{\rho}_0(\bbox{r},\bbox{r}')\delta_{ss'}
+ \breve{\bbox{s}}_0(\bbox{r},\bbox{r}')\cdot \hat{\bbox{\sigma}}_{ss'}\right)\delta_{tt'}\nonumber \\
&+& \tquart\sum_k\left(\delta_{ss'}\breve{\rho}_k(\bbox{r},\bbox{r}')
+ \breve{\bbox{s}}_k(\bbox{r},\bbox{r}')\cdot\hat{\bbox{\sigma}}_{ss'}\right)\hat{\tau}_{tt'}^k,
\label{izo02}
\en
where $k=0,\ 1,\ 2,\ 3$.
The spin-isospin components of the p-h ($\rho_k$, $\bbox{s}_k$) and
p-p
($\breve{\rho}_k$, $\breve{\bbox{s}}_k$)
nonlocal densities are functions of two position vectors $\bbox{r}$ and
$\bbox{r}'$ and have the following symmetry properties that result from
Eqs.~(\ref{herm}) and (\ref{antys}):
\bn
\rho_k(\bbox{r},\bbox{r}')
&=&\rho^{*}_k(\bbox{r}',\bbox{r}), \nonumber \\
\bbox{s}_k(\bbox{r},\bbox{r}')
&=&\bbox{s}^{*}_k(\bbox{r}',\bbox{r}),\label{hermden}
\en
for $k=0,1,2,3$, and:
\bn
\breve{\rho}_k        (\bbox{r},\bbox{r}')
&=& \mp \breve{\rho}_k                (\bbox{r}',\bbox{r}), \nonumber \\
\breve{\bbox{s}}_k    (\bbox{r},\bbox{r}')
&=& \pm \breve{\bbox{s}}_k (\bbox{r}',\bbox{r}), \label{antysden}
\en
where the upper sign is for $k=0$ (isoscalars) and the lower for $k=1,2,3$ (isovectors).

In general, the p-h and  p-p density matrices transform differently
under the single-particle unitary transformations.
However, it is proved in I that for the spatial transformations such as
rotations, space-inversion, etc., the transformation rules for the
p-p density matrix are the same as those for the p-h density matrix.
These rules are obviously the same for all the isospin components.
Therefore, in further discussion of the space symmetries, we shall
omit the accent ``breve" and the index $k$. Also, we shall not take
into account conditions  (\ref{hermden}) and (\ref{antysden}), i.e.,  the
hermiticity   of the p-h densities and  the symmetry or antisymmetry
of the p-p densities. We mention only that the former condition
ensures the reality of all the p-h local densities, whereas the latter
one results in vanishing of either the isoscalar or the isovector p-p
local densities.

Within the local
density approximation,  the
energy functional is built from
the local densities  ($\bbox{r}=\bbox{r}'$) and derivatives thereof.
The exact definitions of all used local
densities are given, e.g., in Refs.\cite{Eng75,Per04}. Here we only provide
schematic definitions that clearly expose  the corresponding spatial properties.
The densities of interest are:
\begin{itemize}
\item zero-order local densities
\begin{itemize}
\item particle or pairing density
\beq
\rho (\bbox{r})=\rho (\bbox{r},\bbox{r}')_{\bbox{r}=\bbox{r}'} \label{r}
\eeq
\item spin density
\beq
\bbox{s}(\bbox{r})=\bbox{s}(\bbox{r},\bbox{r}')_{\bbox{r}=\bbox{r}'} \label{s}
\eeq
\end{itemize}
\item first-order local densities
\begin{itemize}
\item current density
\beq
{\bbox{j}}(\bbox{r})
=\tfrac{1}{2i}\big[ (\bbox{\nabla} - \bbox{\nabla}')
{\rho}(\bbox{r},\bbox{r}')\big]_{\bbox{r}=\bbox{r}'}\label{j}
\eeq
\item spin-current density
\beq
{{\mathsf J}}(\bbox{r})
=\tfrac{1}{2i}\big[ (\bbox{\nabla} - \bbox{\nabla}')\otimes
{\bbox{s}}(\bbox{r},\bbox{r}')\big]_{\bbox{r}=\bbox{r}'}, \label{J}
\eeq
which is decomposed into the spin-divergence (trace of $\mathsf{J}$)
density $J$, the spin-curl (antisymmetric part of $\mathsf{J}$)
density $\bbox{J}$, and the traceless symmetric spin-current
density\footnote{Underlined symbols stand for tensors after symmetrizing and
subtracting the trace, e.g.,
$\underline{\mathsf{J}}_{ab}=\tfrac{1}{2}({\mathsf{J}}_{ab}+{\mathsf{J}}_{ba})
-\tfrac{1}{3}\delta_{ab}\sum_c{\mathsf{J}}_{cc}$.} $\underline{\mathsf{J}}$:
\bn
 J(\bbox{r})
&=&\tfrac{1}{2i}\big[ (\bbox{\nabla} - \bbox{\nabla}')\cdot
{\bbox{s}}(\bbox{r},\bbox{r}')\big]_{\bbox{r}=\bbox{r}'},\label{div}\\
{\bbox{J}}(\bbox{r})
&=&\tfrac{1}{2i}\big[ (\bbox{\nabla} - \bbox{\nabla}')\times
{\bbox{s}}(\bbox{r},\bbox{r}')\big]_{\bbox{r}=\bbox{r}'},\label{curl}\\
\underline{{\mathsf J}}(\bbox{r})
&=&\tfrac{1}{2i}\big[ \underline{(\bbox{\nabla} - \bbox{\nabla}')\otimes
{\bbox{s}}(\bbox{r},\bbox{r}')}\big]_{\bbox{r}=\bbox{r}'} \label{trless}
\en
\end{itemize}
\item second-order local densities
\begin{itemize}
\item kinetic density
\beq
{\tau}(\bbox{r})
=\big[        (\bbox{\nabla}\cdot\bbox{\nabla}')
{\rho}(\bbox{r},\bbox{r}')\big]_{\bbox{r}=\bbox{r}'}\label{t}
\eeq
\item spin-kinetic density
\beq
\bbox{T}(\bbox{r})
=\big[        (\bbox{\nabla}\cdot\bbox{\nabla}')
\bbox{s}(\bbox{r},\bbox{r}')\big]_{\bbox{r}=\bbox{r}'}\label{T}
\eeq
\item spin-tensor density
\beq
{\bbox{F}}(\bbox{r})
=\thalf\big[       (\bbox{\nabla} \!\otimes\!\bbox{\nabla}'
\!+\!\bbox{\nabla}'\!\otimes\!\bbox{\nabla})\!\cdot\!
{\bbox{s}}(\bbox{r},\bbox{r}')\big]_{\bbox{r}=\bbox{r}'}\label{F}.
\eeq
\end{itemize}
\end{itemize}
We confine ourselves to the second-order derivatives as is usually
done. But our analysis can also be extended to higher-order derivatives
of the nonlocal densities that
have recently been considered.\cite{Car08}

\section{Spherical symmetry}\label{sph}

\subsection{Nonlocal and local fields}\label{sphfield}
Let $\bbox{r}$ and $\bbox{r}'$ be two arbitrary, linearly independent
position vectors. Then the vector product
$\bbox{r}\times\bbox{r}'$ is the third linearly independent vector,
and all three form a basis in the three-dimensional space of
positions. The scalar products: $\bbox{r}\cdot\bbox{r}=r^2$,
$\bbox{r}'\cdot\bbox{r}'=r^{\prime 2}$, and $\bbox{r}\cdot\bbox{r}'$
form three independent scalars quadratic in $\bbox{r}$, $\bbox{r}'$.
It is impossible to form a cubic scalar because
$\bbox{r}\cdot (\bbox{r}\times\bbox{r}')=\bm{0}$ and $\bbox{r}'\cdot
(\bbox{r}\times\bbox{r}')=\bm{0}$. Six possible outer products of the
three vectors in question form the following second rank Cartesian
tensors:
\begin{itemize}
\item[-] three quadratic tensors ---
$\bbox{r}\otimes\bbox{r}$, $\bbox{r}'\otimes\bbox{r}'$, and
$\bbox{r}\otimes\bbox{r}'$;\\ the first two tensors are symmetric
and their traces are $r^2$ and $r^{\prime 2}$, respectively;\\ the
vector antisymmetric part of the third tensor  is
$\bbox{r}\times\bbox{r}'$ and its trace is $\bbox{r}\cdot\bbox{r}'$.
\item[-] two traceless cubic tensors ---
$\bbox{r}\otimes(\bbox{r}\times\bbox{r}')$ and
$\bbox{r}'\otimes(\bbox{r}\times\bbox{r}')$ with the vector
antisymmetric parts equal to
$\bbox{r}\times(\bbox{r}\times\bbox{r}')=(\bbox{r}\cdot\bbox{r}')\bbox{r} -r^2\bbox{r}'$ and
$\bbox{r}'\times(\bbox{r}\times\bbox{r}')=-(\bbox{r}\cdot\bbox{r}')\bbox{r}' -r^{\prime 2}\bbox{r}$, respectively.
\item[-] one fourth-order tensor
$(\bbox{r}\times\bbox{r}')\otimes (\bbox{r}\times\bbox{r}')$ which can be expressed as
a linear combination of the quadratic tensors with scalar
coefficients:
\bn
(\bbox{r}\times\bbox{r}')\otimes (\bbox{r}\times\bbox{r}')&=&(\bbox{r}\cdot\bbox{r}')(\bbox{r}\otimes\bbox{r}'+\bbox{r}'\otimes\bbox{r})
-(\bbox{r}\cdot\bbox{r}')^2\mathsf{1}\nonumber \\
&-&r^2(\bbox{r}'\otimes\bbox{r}')-r^{\prime 2}(\bbox{r}\otimes\bbox{r})+r^2r^{\prime 2}\mathsf{1}, \label{syzyg}
\en
where $(\mathsf{1})_{ab}=\delta_{ab}$ ($a,\ b=x,\ y,\ z$) is the unit tensor.
\end{itemize}

Having listed all the independent scalars, vectors, and tensors that can be constructed
from vectors $\bbox{r}$ and $\bbox{r}'$, we are able to give general
expressions for  the nonlocal isotropic fields depending on the two position
vectors. We note that  (i) any scalar field must be an arbitrary function of the independent
scalar functions:
\beq
Q(\bbox{r},\bbox{r}')=q_0(r^2,\bbox{r}\cdot\bbox{r}',r^{\prime 2}); \label{nonlocsc}
\eeq
(ii) any vector field must be a linear combination of $\bbox{r}$, $\bbox{r}'$,
and $\bbox{r}\times\bbox{r}'$ with scalar coefficients:
\beq
\bbox{Q}(\bbox{r},\bbox{r}')=q_{11}(r^2,\bbox{r}\cdot\bbox{r}',r^{\prime 2})\bbox{r}+q_{12}(r^2,\bbox{r}\cdot\bbox{r}',r^{\prime 2})\bbox{r}'
+q_{13}(r^2,\bbox{r}\cdot\bbox{r}',r^{\prime 2})\bbox{r}\times\bbox{r}'; \label{nonlocvec}
\eeq
and (iii) any symmetric traceless tensor field must be a linear combination of the five basic tensors:
\bn
\underline{\mathsf{Q}}(\bbox{r},\bbox{r}')&=&
   q_{21}(r^2,\bbox{r}\cdot\bbox{r}',r^{\prime 2})\underline{\bbox{r}\otimes\bbox{r}}
  +q_{22}(r^2,\bbox{r}\cdot\bbox{r}',r^{\prime 2})\underline{\bbox{r}'\otimes\bbox{r}'}
  +q_{23}(r^2,\bbox{r}\cdot\bbox{r}',r^{\prime 2})\underline{\bbox{r}\otimes\bbox{r}'}
   \nonumber \\
&&+q_{24}(r^2,\bbox{r}\cdot\bbox{r}',r^{\prime 2})\underline{\bbox{r}\otimes(\bbox{r}\times\bbox{r}')}
  +q_{25}(r^2,\bbox{r}\cdot\bbox{r}',r^{\prime 2})\underline{\bbox{r}'\otimes(\bbox{r}\times\bbox{r}')} \nonumber \\
                                          &=&q_{21}(r^2,\bbox{r}\cdot\bbox{r}',r^{\prime 2})(\bbox{r}\otimes\bbox{r}-\tfrac{1}{3}r^2\mathsf{1})
+q_{22}(r^2,\bbox{r}\cdot\bbox{r}',r^{\prime 2})(\bbox{r}'\otimes\bbox{r}'-\tfrac{1}{3}r^{\prime 2}\mathsf{1}) \nonumber \\
&&+q_{23}(r^2,\bbox{r}\cdot\bbox{r}',r^{\prime 2})[\thalf (\bbox{r}\otimes\bbox{r}'+\bbox{r}'\otimes\bbox{r})-\tfrac{1}{3}(\bbox{r}\cdot\bbox{r}')\mathsf{1}]
   \nonumber \\
&&+q_{24}(r^2,\bbox{r}\cdot\bbox{r}',r^{\prime 2})\thalf (\bbox{r}\otimes(\bbox{r}\times\bbox{r}')+(\bbox{r}\times\bbox{r}')\otimes\bbox{r})
   \nonumber \\
&&+q_{25}(r^2,\bbox{r}\cdot\bbox{r}',r^{\prime 2})\thalf (\bbox{r}'\otimes(\bbox{r}\times\bbox{r}')+(\bbox{r}\times\bbox{r}')\otimes\bbox{r}'). \label{nonlocten}
\en
In the expressions above, all $q$'s are arbitrary scalar functions. Scalar fields always have
the positive parity. The parities of vector and tensor fields are, in
general, indefinite. However, since each independent vector or tensor
field does have a definite parity, the vector and tensor fields
of definite parities can be easily defined.

It is readily seen from Eqs.~(\ref{nonlocsc}), (\ref{nonlocvec}), and
(\ref{nonlocten}) that the corresponding local fields, which depend
on one position vector $\bbox{r}=\bbox{r}'$, only take very simple forms (cf.
Appendix A in I):
\bn
Q(\bbox{r})&=&q_0(r^2), \label{locsc}\\
\bbox{Q}(\bbox{r})&=&q_1(r^2)\bbox{r}, \label{locvec}\\
\underline{\mathsf{Q}}(\bbox{r})&=&q_2(r^2)(\bbox{r}\otimes\bbox{r}-\tfrac{1}{3}r^2\mathsf{1}).\label{locten}
\en

\subsection{Nonlocal and local densities}\label{sphden}

\subsubsection{Rotational symmetry SO(3)}\label{so3}

If we assume that the density matrices $\hat{\rho}$
and $\hat{\breve{\rho}}$, Eqs.~(\ref{rho}) and (\ref{rhobreve}), are invariant under
the three-dimensional rotations forming the SO(3) group, it immediately follows from
Eqs.~(\ref{izo01}) and (\ref{izo02}) that the densities of type
$\rho$ are the SO(3) scalars while the densities of type
$\bbox{s}$ are the SO(3) vectors (note that the spin Pauli matrices are the SO(3)
vectors). Therefore, the nonlocal density of
type $\rho$ takes a form of Eq.~(\ref{nonlocsc}):
\beq
\rho (\bbox{r},\bbox{r}')=\varrho_0(r^2,\bbox{r}\cdot\bbox{r}',r^{\prime 2}), \label{nonlocrho}
\eeq
while the nonlocal density of type $\bbox{s}$ has a form of Eq.~(\ref{nonlocvec}):
\beq
\bbox{s}(\bbox{r},\bbox{r}')=
 \varrho_{11}(r^2,\bbox{r}\cdot\bbox{r}',r^{\prime 2})\bbox{r}
+\varrho_{12}(r^2,\bbox{r}\cdot\bbox{r}',r^{\prime 2})\bbox{r}'
+\varrho_{13}(r^2,\bbox{r}\cdot\bbox{r}',r^{\prime 2})\bbox{r}\times\bbox{r}', \label{nonlocs}
\eeq
where $\varrho_0$, $\varrho_{11}$, $\varrho_{12}$, and $\varrho_{13}$ are
arbitrary scalar functions. All local differential densities
(\ref{j})--(\ref{F}) can be calculated by differentiating
Eqs.~(\ref{nonlocrho}) and (\ref{nonlocs}), like it was done in
I. But to establish general forms of  all local densities, it is
sufficient to realize that all of them are local isotropic fields
with  definite transformation rules under  SO(3) rotations. These
rules can be deduced from the definitions (\ref{r})--(\ref{F}). We
see that the local densities $\rho (\bbox{r})$,
 $\tau (\bbox{r})$, and
$J(\bbox{r})$ are scalar fields and all take the form of Eq.~(\ref{locsc}).
Similarly, the densities $\bbox{s}(\bbox{r})$,  $\bbox{j}(\bbox{r})$,  $\bbox{J}(\bbox{r})$,
$\bbox{T}(\bbox{r})$, and  $\bbox{F}(\bbox{r})$ are
the SO(3) vectors; hence, are given by the form of  Eq.~(\ref{locvec}). Finally, the
spin-current density $\underline{\mathsf{J}}(\bbox{r})$ is the
traceless symmetric tensor of the form (\ref{locten}).

\subsubsection{Rotational and mirror symmetry O(3)}\label{o3}

When the density matrices are also invariant  under  mirror reflections, it follows from Eqs.~(\ref{izo01}),
(\ref{izo02}), and (\ref{nonlocrho}) that $\rho (\bbox{r},\bbox{r}')$ should have
positive parity. The spin Pauli matrices form  an O(3)
pseudovector and thus the  nonlocal spin density should be a
pseudovector as well. On the right-hand side of Eq.~(\ref{nonlocs}), only
the last term is a  pseudovector. Therefore, in the case of the O(3)
symmetry, the spin nonlocal density takes the form:
\beq
\bbox{s}(\bbox{r},\bbox{r}')=
\varrho_1''(r^2,\bbox{r}\cdot\bbox{r}',r^{\prime 2}) (\bbox{r}\times\bbox{r}'), \label{nonlocso3}
\eeq
meaning that the local spin density $\bbox{s}(\bbox{r})=0$. It is
impossible to build a pseudovector from one  vector $\bbox{r}$; therefore,
local densities $\bbox{T}(\bbox{r})$ and $\bbox{F}(\bbox{r})$, being
pseudovectors, must vanish, as well as  pseudoscalar
$J(\bbox{r})=0$ and  pseudotensor
$\underline{\mathsf{J}}(\bbox{r})=0$. On the other hand, vectors
$\bbox{j}(\bbox{r})$ and  $\bbox{J}(\bbox{r})$ do not vanish and  take the form (\ref{locvec}).

\section{Axial symmetry}\label{axial}
Let us suppose that rotations and mirror rotations
around one axis (say $z$-axis) are SCSs. This symmetry
group will be denoted as O$^{z\perp}(2)\subset$O(3). It is the direct
product O$^{z\perp}(2)=$S$_z\otimes$SO$^{\perp}(2)$ of the
SO$^{\perp}(2)$ group of rotations around the $z$-axis and the
two-element group S$_z$ consisting of the reflection in the plane
perpendicular to the symmetry axis and the identity. To investigate
the axial symmetry, it is convenient to decompose the position vector
$\bbox{r}$ into the components parallel and perpendicular to the
symmetry axis:
\beq
\bbox{r}=\bbox{r}_z+\bbox{r}_{\perp} , \label{rzp}
\eeq
which have different transformation properties under the
O$^{z\perp}(2)$ transformations. The component $\bbox{r}_{\perp}$ is
a SO$^{\perp}(2)$ vector whereas $\bbox{r}_z$ is not affected by the
SO$^{\perp}(2)$ rotations; hence, it is invariant under SO$^{\perp}(2)$.
On the other hand, $\bbox{r}_z$ changes
its sign under the reflection S$_z$, while $\bbox{r}_{\perp}$ is
S$_z$-invariant.

\subsection{Two-dimensional rotational symmetry SO$^{\perp}(2)$}\label{so2}

If the density matrices $\hat{\rho}$ and
$\hat{\breve{\rho}}$  possess the
SO$^{\perp}(2)$ symmetry, it follows from Eqs.~(\ref{izo01}) and
(\ref{izo02}) that all densities of type $\rho$ should be
SO$^{\perp}(2)$ scalars.
Now it is convenient to decompose densities of type $\bbox{s}$
into components perpendicular and parallel to the symmetry axis:
\beq\label{szp}
\bbox{s}(\bbox{r},\bbox{r}')=\bbox{s}_{\perp}(\bbox{r},\bbox{r}')
                            +\bbox{s}_z(\bbox{r},\bbox{r}') .
\eeq
The perpendicular components
$\bbox{s}_{\perp}(\bbox{r},\bbox{r}')$
are  SO$^{\perp}(2)$ vectors and the parallel components
$\bbox{s}_z(\bbox{r},\bbox{r}')$ are
SO$^{\perp}(2)$ invariants.

The following SO$^{\perp}(2)$
scalars can be constructed from the perpendicular and parallel components
of the
position vectors $\bbox{r}$ and $\bbox{r}'$: $z,\ z',$
the $z$-coordinates of $\bbox{r}_z$ and $\bbox{r}'_z$, respectively,
$r^2_{\perp}=\bbox{r}_{\perp}\cdot\bbox{r}_{\perp},\
\bbox{r}_{\perp}\cdot\bbox{r}'_{\perp},\ r^{\prime2}_{\perp}=\bbox{r}'_{\perp}\cdot\bbox{r}'_{\perp}$,
and $\bbox{r}_z\cdot(\bbox{r}_{\perp}\times\bbox{r}'_{\perp})$ and
$\bbox{r}'_z\cdot(\bbox{r}_{\perp}\times\bbox{r}'_{\perp})=(z'/z)\bbox{r}_z\cdot(\bbox{r}_{\perp}\times\bbox{r}'_{\perp})$.
As we see the two latter scalars are dependent on each other because $\bbox{r}_z\parallel\bbox{r}'_z$.
The square $(\bbox{r}_z\cdot(\bbox{r}_{\perp}\times\bbox{r}'_{\perp}))^2
=z^2(r^2_{\perp}r'^2_{\perp}-(\bbox{r}_{\perp}\cdot\bbox{r}'_{\perp})^2)$ is
dependent on other scalars, namely on $z^2,\ r_{\perp}^2,\ r_{\perp}^{\prime 2}$ and $\bbox{r}_{\perp}\cdot\bbox{r}'_{\perp}$.
Hence, the
nonlocal scalar densities have the form
\beq
\rho (\bbox{r}, \bbox{r}')=\varrho_0(z,z',r^2_{\perp},\bbox{r}_{\perp}\cdot\bbox{r}'_{\perp},r^{\prime 2}_{\perp},
\bbox{r}_z\cdot(\bbox{r}_{\perp}\times\bbox{r}'_{\perp}) ),\label{nonlocsczp}
\eeq
where $\varrho_0$ is an arbitrary scalar function, linear in the last argument. There are three
vectors invariant under SO$^{\perp}(2)$ and parallel to the symmetry
axis, namely $\bbox{r}_z$, $\bbox{r}'_z$, and
$\bbox{r}_{\perp}\times\bbox{r}'_{\perp}$. Consequently, the z-component
of $\bbox{s}(\bbox{r}$, $\bbox{r}')$ is of the form
\bn
\bbox{s}_z(\bbox{r}, \bbox{r}')&=&\varrho_z(z,z',r^2_{\perp},\bbox{r}_{\perp}\cdot\bbox{r}'_{\perp},r^{\prime 2}_{\perp},
\bbox{r}_z\cdot(\bbox{r}_{\perp}\times\bbox{r}'_{\perp}))\bbox{r}_z \nonumber \\
&+&\varrho_z'(z,z',r^2_{\perp},\bbox{r}_{\perp}\cdot\bbox{r}'_{\perp},r^{\prime 2}_{\perp},
\bbox{r}_z\cdot(\bbox{r}_{\perp}\times\bbox{r}'_{\perp}))\bbox{r}'_z \nonumber \\
&+&\varphi_z(z,z',r^2_{\perp},\bbox{r}_{\perp}\cdot\bbox{r}'_{\perp},r^{\prime 2}_{\perp},
\bbox{r}_z\cdot(\bbox{r}_{\perp}\times\bbox{r}'_{\perp}))(\bbox{r}_{\perp}\times\bbox{r}'_{\perp}),
\label{nonlocsz}
\en
where $\varrho_z$, $\varrho_z'$, and $\varphi_z$ are scalar functions.
There exist  four independent SO$^{\perp}(2)$ vectors that lie in the
plane perpendicular to the symmetry axis: $\bbox{r}_{\perp}$,
$\bbox{r}'_{\perp}$, $\bbox{r}_z\times\bbox{r}_{\perp}$ and
$\bbox{r}'_z\times\bbox{r}'_{\perp}$. Hence, the SO$^{\perp}(2)$
vector component of the nonlocal spin density takes the
form:
\bn
\bbox{s}_{\perp}(\bbox{r}, \bbox{r}')&=&\varrho_{\perp}(z,z',r^2_{\perp},\bbox{r}_{\perp}\cdot\bbox{r}'_{\perp},r^{\prime 2}_{\perp},
\bbox{r}_z\cdot(\bbox{r}_{\perp}\times\bbox{r}'_{\perp}))\bbox{r}_{\perp} \nonumber \\
&&+\varrho_{\perp}'(z,z',r^2_{\perp},\bbox{r}_{\perp}\cdot\bbox{r}'_{\perp},r^{\prime 2}_{\perp},
\bbox{r}_z\cdot(\bbox{r}_{\perp}\times\bbox{r}'_{\perp}))\bbox{r}'_{\perp} \nonumber \\
&&+\varphi_{\perp}(z,z',r^2_{\perp},\bbox{r}_{\perp}\cdot\bbox{r}'_{\perp},r^{\prime 2}_{\perp},
\bbox{r}_z\cdot(\bbox{r}_{\perp}\times\bbox{r}'_{\perp}))(\bbox{r}_z\times\bbox{r}_{\perp}) \nonumber \\
&&+\varphi_{\perp}'(z,z',r^2_{\perp},\bbox{r}_{\perp}\cdot\bbox{r}'_{\perp},r^{\prime 2}_{\perp},
\bbox{r}_z\cdot(\bbox{r}_{\perp}\times\bbox{r}'_{\perp}))(\bbox{r}'_z\times\bbox{r}'_{\perp}),
\label{nonlocsp}
\en
with $\varrho_{\perp}$, $\varrho_{\perp}'$, $\varphi_{\perp}$, and
$\varphi_{\perp}'$ being scalar functions.
General forms of the local zero-order densities are obtained
 from Eqs.~(\ref{nonlocsczp}), (\ref{nonlocsz}), and
(\ref{nonlocsp}) by putting  $\bbox{r}_z =\bbox{r}'_z$ and
$\bbox{r}_{\perp}=\bbox{r}'_{\perp}$. This gives:
\bn
\rho (\bbox{r})&=&\rho_0(z,r_{\perp}), \label{locsczp}\\
\bbox{s}_z(\bbox{r})&=&\rho_z(z,r_{\perp})\bbox{r}_z, \label{locsz}\\
\bbox{s}_{\perp}(\bbox{r})&=& \rho_{\perp}(z,r_{\perp})\bbox{r}_{\perp}+\phi_{\perp}(z,r_{\perp})(\bbox{r}_z\times\bbox{r}_{\perp}),\label{locsp}
\en
where $\rho_0$, $\rho_z$, $\rho_{\perp}$, and $\phi_{\perp}$ are
arbitrary scalar functions of $z$ and
$r_{\perp}=\sqrt{r^2_{\perp}}$. The general form of the remaining
scalar densities $\tau (\bbox{r})$ and $J(\bbox{r})$  is the same as that of
Eq.~(\ref{locsczp}).

The components of the differential operator,
\beq
\bbox{\nabla}=\bbox{\nabla}_z+\bbox{\nabla}_{\perp} , \label{nablazp}
\eeq
have the same transformation properties under  SO$^{\perp}(2)$
rotations as the corresponding components of the position
vector (\ref{rzp}). Hence, the densities $\bbox{j}(\bbox{r})$,
$\bbox{J}(\bbox{r})$, $\bbox{T}(\bbox{r})$, and $\bbox{F}(\bbox{r})$
 have the components parallel to the $z$-axis that are invariant
under  SO$^{\perp}(2)$ and the SO$^{\perp}(2)$ vector components
that are
perpendicular to the symmetry axis. Therefore, they all  take general forms
given by  Eqs.~(\ref{locsz}) and
(\ref{locsp}).

It follows from the definitions (\ref{J}),
(\ref{trless}), and (\ref{nablazp}) that the components
$(\underline{\mathsf{J}})_{az}$ ($a=x,\ y$) of the symmetric
traceless spin-current density form the SO$^{\perp}(2)$ vector while
the components $(\underline{\mathsf{J}})_{ab}$ ($a,\ b=x,\ y$) form
the SO$^{\perp}(2)$ symmetric traceless tensor. The following four
symmetric traceless tensors can be formed with vectors $\bbox{r}_z$
and $\bbox{r}_{\perp}$:
\bn
\underline{\bbox{r}_z\otimes\bbox{r}_{\perp}}&=&\tfrac{1}{2}(\bbox{r}_z\otimes\bbox{r}_{\perp}+\bbox{r}_{\perp}\otimes\bbox{r}_z), \label{tzp} \\
\underline{\bbox{r}_z\otimes(\bbox{r}_z\times\bbox{r}_{\perp}})&=&
\tfrac{1}{2}(\bbox{r}_z\otimes(\bbox{r}_z\times\bbox{r}_{\perp})+(\bbox{r}_z\times\bbox{r}_{\perp})\otimes\bbox{r}_z), \label{tzzp} \\
\underline{\bbox{r}_{\perp}\otimes\bbox{r}_{\perp}}&=&\bbox{r}_{\perp}\otimes\bbox{r}_{\perp}-\tfrac{1}{2}r^2_{\perp}\mathsf{1}_{\perp}, \label{tpp}\\
\underline{\bbox{r}_{\perp}\otimes(\bbox{r}_z\times\bbox{r}_{\perp}})&=&
\tfrac{1}{2}(\bbox{r}_{\perp}\otimes(\bbox{r}_z\times\bbox{r}_{\perp})+(\bbox{r}_z\times\bbox{r}_{\perp})\otimes\bbox{r}_{\perp}), \label{tpzp}
\en
where $(\mathsf{1}_{\perp})_{ab}=(1-\delta_{az})\delta_{ab}$ ($a,\
b=x,\ y,\ z$) is the unit tensor in the plane perpendicular to the
symmetry axis. Two  tensors (\ref{tzp}) and
(\ref{tzzp}) transform under  SO$^{\perp}(2)$  like
vectors perpendicular to the $z$-axis. The tensors
(\ref{tpp}) and (\ref{tpzp}) are the  SO$^{\perp}(2)$ tensors. Consequently, the general form of the symmetric traceless spin-current
density is:
\bn\label{trlesszp}
\underline{\mathsf{J}}(\bbox{r})&=&\theta_{z\perp}(z,r_{\perp})\underline{\bbox{r}_z\otimes\bbox{r}_{\perp}}
+\theta_{zz\perp}(z,r_{\perp})\underline{\bbox{r}_z\otimes(\bbox{r}_z\times\bbox{r}_{\perp})} \nonumber \\
&+&\theta_{\perp\perp}(z,r_{\perp})\underline{\bbox{r}_{\perp}\otimes\bbox{r}_{\perp}}
+\theta_{\perp z\perp}(z,r_{\perp})\underline{\bbox{r}_{\perp}\otimes(\bbox{r}_z\times\bbox{r}_{\perp})},
\en
where $\theta_{z\perp}$, $\theta_{zz\perp}$, $\theta_{\perp\perp}$,
and $\theta_{\perp z\perp}$ are scalar functions. We note in passing
that in the case of the axial symmetry, all local densities
formally look like the spherical symmetric nonlocal fields
(\ref{nonlocsc}), (\ref{nonlocvec}), and (\ref{nonlocten}) with
$\bbox{r}=\bbox{r}_z$ and $\bbox{r}'=\bbox{r}_{\perp}$, provided that
$\bbox{r}_z\cdot\bbox{r}_{\perp}=0$.

\subsection{Two-dimensional rotational and mirror symmetry O$^{z\perp}(2)$}\label{o2}

It follows from
the S$_z$ invariance of the density matrices (\ref{rho}) and
(\ref{rhobreve}) that $\rho (\bbox{r},\bbox{r}')$ and
$\bbox{s}_z(\bbox{r},\bbox{r}')$ do not change  signs under
S$_z$, while $\bbox{s}_{\perp}(\bbox{r},\bbox{r}')$ does. All scalar
functions invariant under S$_z$ depend on $z$, $z'$ only through
$z^2=\bbox{r}_z\cdot\bbox{r}_z$, $z^{\prime
2}=\bbox{r}'_z\cdot\bbox{r}'_z$, and $zz'=\bbox{r}_z\cdot\bbox{r}'_z$.
For instance, Eq.~(\ref{nonlocsczp}) for the nonlocal scalar density
now takes the form
\beq
\rho (\bbox{r}, \bbox{r}')=\varrho_0(z^2,zz',z'^2,r^2_{\perp},\bbox{r}_{\perp}\cdot\bbox{r}'_{\perp},r^{\prime 2}_{\perp}). \label{nonlocsco2}
\eeq
There exist  two pseudoscalars formed from
$\bbox{r}_z$, $\bbox{r}'_z$, $\bbox{r}_{\perp}$, $\bbox{r}'_{\perp}$, namely
$\bbox{r}_z\cdot(\bbox{r}_{\perp}\times\bbox{r}'_{\perp})$ and
$\bbox{r}'_z\cdot(\bbox{r}_{\perp}\times\bbox{r}'_{\perp})=(zz'/z^2)\bbox{r}_z\cdot(\bbox{r}_{\perp}\times\bbox{r}'_{\perp})$
equal to each other up to the scalar factor $zz'/z^2$.
The square $(\bbox{r}_z\cdot(\bbox{r}_{\perp}\times\bbox{r}'_{\perp}))^2=z^2(r^2_{\perp}r'^2_{\perp}-(\bbox{r}_{\perp}\cdot\bbox{r}'_{\perp})^2)$ is,
of course, a scalar.

To fulfill the transformation rules, the general forms of Eqs.~(\ref{nonlocsz}) and (\ref{nonlocsp}) of the components of the nonlocal spin density
should be modified in the following way:
\bn
\bbox{s}_z(\bbox{r}, \bbox{r}')&=&
(\bbox{r}_z\cdot(\bbox{r}_{\perp}\times\bbox{r}'_{\perp}))\varrho_z(z^2,zz',z'^2,r^2_{\perp},\bbox{r}_{\perp}\cdot\bbox{r}'_{\perp},r^{\prime 2}_{\perp})
\bbox{r}_z \nonumber \\
&&+(\bbox{r}'_z\cdot(\bbox{r}_{\perp}\times\bbox{r}'_{\perp}))
\varrho_z'(z^2,zz',z'^2,r^2_{\perp},\bbox{r}_{\perp}\cdot\bbox{r}'_{\perp},r^{\prime 2}_{\perp})\bbox{r}'_z \nonumber \\
&&+\varphi_z(z^2,zz',z'^2,r^2_{\perp},\bbox{r}_{\perp}\cdot\bbox{r}'_{\perp},r^{\prime 2}_{\perp})(\bbox{r}_{\perp}\times\bbox{r}'_{\perp}),
\label{nonlocszo2}
\en
and
\bn
\bbox{s}_{\perp}(\bbox{r}, \bbox{r}')&=&(\bbox{r}_z\cdot(\bbox{r}_{\perp}\times\bbox{r}'_{\perp}))
\varrho_{\perp}(z^2,zz',z'^2,r^2_{\perp},\bbox{r}_{\perp}\cdot\bbox{r}'_{\perp},r^{\prime 2}_{\perp})\bbox{r}_{\perp} \nonumber \\
&&+(\bbox{r}'_z\cdot(\bbox{r}_{\perp}\times\bbox{r}'_{\perp}))
\varrho_{\perp}'(z^2,zz',z'^2,r^2_{\perp},\bbox{r}_{\perp}\cdot\bbox{r}'_{\perp},r^{\prime 2}_{\perp})\bbox{r}'_{\perp} \nonumber \\
&&+\varphi_{\perp}(z^2,zz',z'^2,r^2_{\perp},\bbox{r}_{\perp}\cdot\bbox{r}'_{\perp},r^{\prime 2}_{\perp})(\bbox{r}_z\times\bbox{r}_{\perp}) \nonumber \\
&&+\varphi_{\perp}'(z^2,zz',z'^2,r^2_{\perp},\bbox{r}_{\perp}\cdot\bbox{r}'_{\perp},r^{\prime 2}_{\perp})(\bbox{r}'_z\times\bbox{r}'_{\perp}).
\label{nonlocspo2}
\en

It follows from Eqs.~(\ref{nonlocsco2}), (\ref{nonlocszo2}), and
(\ref{nonlocspo2}) that the local zero-order
densities for $\bbox{r}_z =\bbox{r}'_z$ and
$\bbox{r}_{\perp}=\bbox{r}'_{\perp}$ can be written in the general form:
\bn
\rho (\bbox{r})&=&\rho_0(z^2,r^2_{\perp}), \label{locsco2}\\
\bbox{s}_z(\bbox{r})&=&0, \label{locszo2}\\
\bbox{s}_{\perp}(\bbox{r})&=& \phi_{\perp}(z^2,r^2_{\perp})(\bbox{r}_z\times\bbox{r}_{\perp}).\label{locspo2}
\en
The local kinetic density $\tau (\bbox{r})$ is of the form
(\ref{locsco2}) too. On the other hand, the pseudoscalar  density
$J(\bbox{r})$ vanishes. The
densities $\bbox{T}(\bbox{r})$ and $\bbox{F}(\bbox{r})$,
are  pseudovectors; hence,  they take the form (\ref{locspo2}).
On the other hand, vectors  $\bbox{j}(\bbox{r})$
and $\bbox{J}(\bbox{r})$ are linear
combinations of the components of the position vector:
\beq
\bbox{j}(\bbox{r})=\iota_z(z^2,r^2_{\perp})\bbox{r}_z+\iota_{\perp}(z^2,r^2_{\perp})\bbox{r}_{\perp}. \label{jo2}
\eeq
The spin-curl $\bbox{J}(\bbox{r})$ takes a similar form to that of
(\ref{jo2}). Finally, $\underline{\mathsf{J}}(\bbox{r})$ is a
pseudotensor. Therefore, as follows from (\ref{locten}), its general
form is given by:
\bn\label{trlesso2}
\underline{\mathsf{J}}(\bbox{r})&=&\theta_{zz\perp}(z^2,r^2_{\perp})\underline{\bbox{r}_z\otimes(\bbox{r}_z\times\bbox{r}_{\perp})}\nonumber \\
&+&\theta_{\perp z\perp}(z^2,r^2_{\perp})\underline{\bbox{r}_{\perp}\otimes(\bbox{r}_z\times\bbox{r}_{\perp})}.
\en

\section{Summary}\label{sum}

In the DFT, for both theoretical and
practical reasons, it is important to know what   general forms of
densities are which obey SCS of interest. In the case of the space symmetries, such general forms can
be established by means of methods of constructing the isotropic
tensor fields.

For the spherical symmetry, the local densities are the isotropic
scalar, vector, or (the second rank) tensor fields, depending on the
position vector $\bbox{r}$. The form of an isotropic field with given
rank is unique and determined through one arbitrary scalar function.
In particular, the parity of the field is unique for a given rank.
Pseudoscalar, pseudovector, and pseudotensor fields do not exist. This
is why in the case of the rotational and mirror symmetry, the
pseudoscalar (spin-divergence), pseudovector (spin, spin-kinetic and
tensor-kinetic), and pseudotensor (symmetric spin-current) local
densities vanish.

For the axial symmetry, the local densities are isotropic fields
depending on two components of the position vector: $\bbox{r}_z$ and
$\bbox{r}_{\perp}$.  The case of SO$^{\perp}(2)$ is interesting as it allows us to better understand the
assumption of  the isotropy of a field. Indeed, in this case it
might seem that the local densities are fields, depending on the
SO$^{\perp}(2)$ vector $\bbox{r}_{\perp}$. However, such fields are
not isotropic. There is another vector, $\bbox{r}_z$,
which plays the role of a material vector fixed by the direction of the
symmetry axis of the system.

\section{Appendix: The Generalized Cayley-Hamilton Theorem}\label{GCHT}

The original Cayley-Hamilton Theorem states that every square matrix satisfies
its own characteristic equation (cf. e.g. \cite{Tur46,Bir65}). It immediately follows from
the theorem that the second rank SO(3) Cartesian tensor $\mathsf{q}$ satisfies the equation
\beq\label{CH}
-\mathsf{q}^3+q_1\mathsf{q}^2-q_2\mathsf{q}+q_3=0,
\eeq
where $q_1$ is the trace, $q_2$ is the sum of the principal subdeterminants,
and $q_3$ is the determinant of $\mathsf{q}$. Hence, a second rank tensor field $\mathsf{Q}(\mathsf{q})$
being the power series of the tensor $\mathsf{q}$
\beq\label{powser}
\mathsf{Q}(\mathsf{q})=c_0\mathsf{1}+c_1\mathsf{q}+c_2\mathsf{q}^2+c_3\mathsf{q}^3+\dots
\eeq
can be sumed up to the form:
\beq\label{sfq}
\mathsf{Q}(\mathsf{q})=\rho_0(q_1,q_2,q_3)\mathsf{1}+\rho_1(q_1,q_2,q_3)\mathsf{q}
+\rho_2(q_1,q_2,q_3)\mathsf{q}^2,
\eeq
where $\rho_1$, $\rho_2$ and $\rho_3$ are functions of scalars $q_1$, $q_2$ and $q_3$.
The power series (\ref{powser}) is an isotropic function of $\mathsf{q}$ because it does not
contain any other tensor.\cite{Tru52} For a symmetric traceless $(q_1=0)$
tensor $\underline{\mathsf{Q}}$ (a quadrupole tensor)
Eq.~(\ref{sfq}) takes the form:
\beq\label{strq}
\underline{\mathsf{Q}}(\mathsf{q})=\varrho_1(q_2,q_3)\underline{\mathsf{q}}
+\varrho_2(q_2,q_3)\underline{\mathsf{q}^2}
\eeq
with two functions $\varrho_1$ and $\varrho_2$ of the two scalars $q_2$ and $q_3$.
Eq.~(\ref{strq}) is a direct consequence of the Cayley-Hamilton Theorem and thus
can be called the Cayley-Hamilton theorem for the quadrupole tensors. It can be generalized
for the isotropic tensor field
$Q^{(L)}(q^{(\lambda )},q^{(\lambda^{\prime})},\dots )$ with an arbitrary multipolarity $L$
being a function of one or a few spherical tensors $q^{(\lambda )}$, $q^{(\lambda^{\prime})},\ \dots$
of  ranks $\lambda$, $\lambda^{\prime},\ \dots$, respectively.
The Generalized Cayley-Hamilton Theorem has the following general form:
\beq\label{GCH}
Q^{(L)}(q^{(\lambda )},q^{(\lambda^{\prime})},\dots )=\sum_{k=1}^{k(L,\lambda ,\lambda^{\prime},\dots )}
R_k(q)T^{(L)}_k(q^{(\lambda )},q^{(\lambda^{\prime})},\dots ),
\eeq
where $q$ stands for the set of independent scalars,  $T^{(L)}_k$ are some definite fundamental tensors, all
constructed from
$q^{(\lambda )}$, $q^{(\lambda^{\prime})},\ \dots$, and $R_k$ are arbitrary scalar functions.
The number $k(L,\lambda ,\lambda^{\prime},\dots )$
depends on the ranks of the all  involved tensors. To find the number and the forms of
the fundamental tensors in a general case may appear to be a difficult task. A systematic
method of constructing them in the case of the tensor fields depending on one tensor are presented
in Refs.\cite{Roh09,Pro09}
Eqs.~(\ref{nonlocsc})--(\ref{nonlocten}) are examples of the GCH theorem
for $\lambda =\lambda^{\prime}=1$ and $L=0,\ 1,\ 2$, whereas Eqs.~(\ref{locsc})--(\ref{locten}) ---
for $\lambda =1$ and $L=0,\ 1,\ 2$. In the nuclear collective model the GCH theorem was used for $\lambda =2$
and $\lambda =3$.\cite{Eis87,Roh78}

\section*{Acknowledgements}

This work was supported in part by the Polish Ministry of Science
under Contract No.~N~N202~328234, by the Academy of Finland and
University of Jyv\"askyl\"a within the FIDIPRO programme, and by the
U.S.\ Department of Energy under Contract Nos.\ DE-FC02-07ER41457
(UNEDF SciDAC Collaboration), DE-FG02-96ER40963 (University of
Tennessee), and DE-FG05-87ER40361 (Joint Institute for
Heavy Ion Research).

\end{document}